\documentclass[%
  twocolumn,
 showpacs,
 showkeys,
 preprintnumbers,
 amsmath,amssymb,
 aps,
  pra,
  longbibliography,
 ]{revtex4-1}

\usepackage[dvipsnames]{xcolor}

\usepackage{mathptmx}

\usepackage{amssymb,amsthm,amsmath}

\usepackage{tikz}
\usetikzlibrary{calc}

\usepackage{graphicx}
\usepackage{url}

\usepackage{hyperref}
\hypersetup{
    colorlinks=true, 
    linktoc=all,     
    linkcolor=blue,  
    citecolor=blue,
    filecolor=red,
    urlcolor=blue
}

\usepackage{xcolor}

\begin{document}

\title{Theology and Metaphysics in Sombre, Scientific Times}

\author{Karl Svozil}
\email{svozil@tuwien.ac.at}
\homepage{http://tph.tuwien.ac.at/~svozil}

\affiliation{Institute for Theoretical Physics,
Vienna  University of Technology,
Wiedner Hauptstrasse 8-10/136,
1040 Vienna,  Austria}

\date{\today}

\begin{abstract}
In view of the sobering findings of science, theology and to a lesser degree metaphysics is confronted with a humiliating loss, and a need for reinterpretation, of allegories and narratives which have served as guidance to the perplexed for millennia.
Future revolutions of world perception might include the emergence of consciousness and superhuman artificial intelligence from universal computation, extensive virtual reality simulations, the persistence of claims of irreducible chance in the Universe, as well as contacts with alien species and the abundance of inhabited planets.
As tragic and as discomforting as this might be perceived for the religious orthodoxy and by individual believers, a theology guided by science may lead us to a better and more adequate understanding of our existence.
The {\it post factum} theological options are plentiful. These include dualistic scenarios, as well as (to quote Kelly James Clark), a curling or bowling deity, that is, {\em creatio continua}, or {\em ex nihilo}.
These might be grounded in, or corroborated by the metaphysical enigma of existence, which appears to be immune and robust with respect to the aforementioned challenges of science.
\end{abstract}


\maketitle

\onecolumngrid

{
\it
\begin{flushright}
Inspired by and dedicated to \\
Nicholas of Cusa (aka Nicolaus Cusanus)
\end{flushright}
}

\tableofcontents

\newpage

\twocolumngrid

\section{Humiliating aspects of science}

For the individual the Renaissance -- morphing into
Enlightenment  and the scientific revolution -- brought about an ever increasing
quality of life, wealth and abundance of resources due to advances in technology, productivity and trades.
This transition was by no means smooth, as the tragic attempts to silence Giordano Bruno
(who refused and got burned by my church),
Galileo Galilei and Baruch Spinoza document.
Eventually, these improved material conditions contributed towards more
liberal social and political situations.
The open access dissemination of scientific findings
resulted in a corrosion of ancient suspicions and consents, and also in the erosion of
many theological narratives and beliefs; so much so that
even their reinterpretation as allegories~\cite{Maimonides1,Maimonides2} remains precarious.
Many contemporaries live their secular lifes ``as if'' -- that is, fapp: for all practical purposes~\cite{bell-a} --
God is totally superfluous, and maybe a construction of the human mind, but otherwise nonexistent.
This motive is reflected in
{\em divine transcendence} radically denying any kind of {\em ``immanent holiness,''} and by postulating
a total disentanglement between God and the universe; thereby
making any talk of divine intention or intervention
[e.g., Clark's {\em God--as--Curler ({\it versus} Bowler)} metaphor~\cite{Clark-2017-GodAsCurler}]
 -- also duties such as the {\it mitzvot} and adherence to the laws of {\it halakha}~\cite{Soloveitchik-1994} --
a product of one's own imagination; essentially an idolatry~\cite{Statman-2005}.

Socially and politically the disentanglement of the age-old symbiosis between
theocracies and secular rulers (allegedly of God's grace),
resulted in profane societies.
Alas secularism is not grounded in the absolute but is intrinsically means relative.
Therefore, it constantly challenges --
sometimes in violent, misguided and individually detrimental outbursts of revolutions --
the appropriation of resources, as well as the ethics previously dominating religions came packaged with.

On a personal level the individual human mind, ``has had to endure
outrages against its naive self-love''~\cite[Lecture XVIII]{Freud-itpe}
(the German word Freud used is {\em Kr\"ankungen}~\cite[Lecture XVIII]{Freud-itp}),
accompanied by
affronts and traumatic humiliations for their narcissistic perception of grandiosity, and a loss of purpose.
These involved the abandoning of Earth as the center of the universe,
as well as the theory of evolution~\cite{duBois-Reymond-DaC-1883,Horgan-2015},
essentially claiming that humans are on a continuum with, and in many respects not so far away from, plants and animals.
Some recent developments include the comparison of the human and monkey genomes,which has opened up possibilities
to search, locate and identify the
``jewels of the our genome~$\ldots$ underlying the evolutionarily unique capacities of the human brain''~\cite{Sikela-Jewels-2006};
and subsequently the creation of transgenic monkeys with improved short-term memory and reaction time
as hints of improved cognition capacities~\cite{Huang-2019}.
However, this is no reason for grief, as such disillusionments have been more than compensated by gains in individual dignity,
loss of fear, and self-determination: very few would seriously consider going back to the old days of ``blissfull ignorance.''
But this lack of alternatives may even worsen the desperation.

Some of the most sobering thoughts, inspired by this new perception of the universe and
amounting to nihilism can be found in
Schopenhauer's {\em ``The World as Will and Representation''}~\cite[Chapter~1]{schopenhauer-dwawuv-VII};
as well as in
the first paragraph of the young Nietzsche's {\it opus posthumus}
{\em ``On Truth and Lie in an Extra-Moral Sense''}  --
I leave it to the reader to contemplate the depth of melancholy and despair
which overcame this restless troubled mind~\cite{Nietzsche-WahrheitLuege,NietzscheReader}:
{\em ``and yet he still would not have adequately illustrated how miserable, how shadowy and
transient, how aimless and arbitrary the human intellect looks within nature.''}

\section{Humiliations on the horizon}

In what follows I shall mention a few humiliations which are yet to come, and can be expected during our pursuit
further down the scientific lane.
Presently these possibilities are highly speculative, but they are no ``unknown unknowns''~\cite{Rumsfeld-2001}
as they appear on today's horizon of imagination.

I will then argue that at least one metaphysical position remains immune to all such affronts:
that of the enigma of existence, the problem of why there exists something rather than nothing.

\subsection{``Artificial'' (machine) intelligence}
\subsubsection{Emergent consciousness through computational processes}

Already Alan Turing was pondering the possibility that ``artificial'' intelligence capable of higher perceptual
functions such as self-reflection and consciousness
could emerge from {\it ``paper machines}~\cite[p.~34]{Turing-Intelligent_Machinery} (aka {\em universal Turing machines}):
a {\em ``man provided with paper, pencil, and rubber, and subject to strict discipline is
in effect a universal machine.''}
Of course, no ``manly capacity'' is required for its execution;
the requirements imposed on a suitable mechanistic agent (such as a computer processing unit) have been formalized
by the term partial $\mu$--recursive functions
formed by three functional types and three operator types: the constant, the successor and the projection functions,
as well as the composition, the primitive recursion and the minimization operator $\mu$~\cite{Kleene1936,kleene-52}.

Rather than presenting a direct argument for the emergence of ``artificial'' intelligence
Turing concentrated on the {\em imitation game}~\cite{turing:50} of whether or not a paper machine can be built which,
for all practical purposes, in its transactional behavior cannot be discriminated from a human.
Turing was positive about this~\cite{Turing-1996};
and the latest instantiations of agents such as {\em Google Duplex}~\cite{google-duplex}
arguable~\cite{Kosoff-VanityFair-2018,Temperton-Wired-2018}
seem to be able to achieve this goal -- at least for simple dialogues related to logistic tasks.

``Artificial'' intelligence, and, in particular, consciousness,
might be perceived as an emerging function of universal Turing computability.
For the sake of extending Turing's considerations let us formulate the following
hypothesis, and call it the {\em second Turing (hypo)thesis}:
{\em ``given enough time and space (computational capacity), every paper (aka universal Turing) machine
will eventually develop intelligence.''}
Let us call the {\em strong second Turing (hypo)thesis} the following statement:
{\em ``given enough time and space (computational capacity), every paper (aka universal Turing) machine
will eventually not only develop intelligence but also consciousness.''}
In particular, it will pass the Turing's imitation game test,
as well as Greenberger's Genesis test~\cite{greenberger-testint}:
it breaks super rules it was supposed to obey.

\subsubsection{``Artificial'' intelligence outsmarting humans}

Already Turing speculated~\cite[p.~259]{Turing-1996} {\em ``that once the
machine thinking method had started, it would not take long to outstrip
our feeble powers.''}
Indeed, one could speculate that, just like Moore's law was valid until recently,
unbounded machine thinking will reach and go beyond human cognition in a geometric progression, soon ending in a ``singularity'' of almost infinite cognitive capacity;
very much like compound interest is at the heart of the {M}atthew effect~\cite{merton-68}.
Thereby it is not necessary to start from superhuman capacity~\cite{Yudkowsky-96-staring-singularity}
because any nonzero capacity of self-improvement
(by self-reproduction) is sufficient; it would just take ``a little longer.''
Indeed there seems to be no principle preventing intelligence beyond the human capacities~\cite{Bostrom-2014,Bostrom-2017-pdAI}.

In very specific areas this has been achieved already: arguably the strongest {\em chess}
as well as {\em Go} players in history are paper machines.
The latter ones~\cite{AlphaGo-2018} {\em ``played a handful of highly inventive winning moves,
several of which were so surprising they overturned hundreds of years of received wisdom.''}
Recent developments point to directions of autonomous, reflexive learning~\cite{Silver-2017}:
{\em ``reinforcement learning systems
are trained from their own experience, in principle allowing them to
exceed human capabilities, and to operate in domains where human
expertise is lacking.''}
{\em Transfer learning} is based on the idea that training a machine on certain tasks,  thereafter
followed by the elimination of certain layers and components while maintaining ``deeper'' levels of training,
is another method of optimizing performance on totally new tasks.

It can be expected that universal Turing (paper) machines will achieve saturation in intelligence tests way beyond
human capacities, reaching ceiling scores at tests designed to measure human intelligence quotients.
Therefore new schemes will have to be designed to properly measure the super-human capacities of paper machines.

Whether and how human ``fleshware'' will survive an ``artificial'' intelligence supremacy remains to be seen.
Listening to the philosophical, metaphysical, and theological rants of paper machines exceeding human capacities  will be fascinating;
let alone their findings in science and technology~\cite{2016-AlexanderKarlSvozil-ml}.
It is not unlikely that future space probes from Earth -- or alternatively,
close encounters with alien crafts from exoplanets and outer space -- might be populated by such entities,
because it is not too unreasonable to speculate that ``artificial'' intelligent machinery will have developed before interstellar space travel.

Of course, it might be tempting -- and, indeed, this cannot be excluded at the moment --
to maintain that paper machines will never develop consciousness and ``a soul'' in a human sense,
that the cognitive capacities of ``artificial'' intelligence
on the one hand and consciousness on the other hand
must be considered as being completely distinct and not covarying;
even if a machine would pass all conceivable tests and convinces humans of their possessing consciousness.
The ``quality'' of a human self might be very different from any consciousness a paper machine might develop;
just as we pretend this to be true for the cognitive capacities of plants or animals
(otherwise we should consider ourself murderers).
To these objections one may respond that the same criticism may be raised against the existence
of other humans -- after all we may be living in a ``Cartesian prison''~\cite[Meditation~1.12]{descartes-meditation} -- a virtual reality inhabited by
just one consciousness, namely ``us''; that is, you (exclusive) or me.
Another immediate objection might be that the human deficiency in chess and go is akin to trying to compare human performance in say, marathon, with a car.
After all, special purpose machines were developed to outperform humans in a variety of disciplines.
Yet I am inclined to believe that some general cognitive supremacy ``feels'' very differently from motoric, low-complex mechanistic advantages,
or even gaming excesses.

Suppose for a moment that the strong second Turing (hypo)thesis turns out to be correct,
and cognitive capacities and consciousness tend to be covarying.
In such a scenario it might still be
viable to ask whether it might be possible to utilize the cognitive superiority of paper machines
while at the same time preventing them from becoming conscious.
However, one may strongly doubt that in general this is possible,
because restrictions on their processing capacities --
such as the impeding and blocking the implementation of one or more necessary functions and operations --
might strongly diminish their performance and usefulness in other areas.
By definition, as long as an implementation turns out to be Turing universal, there is not way of excluding certain
(computable) functionalities.
Also, it would be impossible to prevent in principle the realization of full implementations of paper machines;
everywhere and always.

\subsection{Extraterrestrial, alien life}

Speculations about the emergence of ``artificial'' intelligence through paper machines are related to
suppositions that also extraterrestrial forms of conscious ``alien'' life have developed.
Indeed one may consider the latter as a corollary of the former:
because the same universe which enabled the material realization of the functions and operators forming paper machines on one planet -- for instance, on Earth --
allows other realizations of paper machines on different planets as well.

So, one may ask~\cite{Jones-1985,Gray-2015}, ``where is everybody?''
Various official~\cite{Condon-report,Condign} and inofficial~\cite{COMETA} reports emphasize that occasionally observed
unidentified flying objects (UFOs) are most likely unidentified aerial phenomena (UAPs) of terrestrial origin.
In any case, if they are hypothetically and speculatively interpreted as alien crafts they are perceived as presenting no threat to the respective national defense~\cite{Wired-UFO}.
There may be a lot of options why no direct and commonly acknowledged contact has happened so far;
the most prominent being the {\em zoo hypothesis}~\cite{Ball1973347}: the potential aliens see no advantage in a mutual exchange and
observe us with a quasi-ethnological interest.

Of course, there exist less benign motives for extraterrestrial aliens for not contacting us;
in particular, if Earth is located within a galactic hypercivilization~\cite{Gato-Rivera-2005}.
An official exchange with supposedly technologically advanced alien civilizations could be perceived as a lose-lose transaction for both sides.
Because from the point of view of the primitive culture (``us'') and their indigenous political, social, scientific and religious institutions,
acknowledgment of some pervasive scientific and technological superiority might be accompanied by a widespread dissolution of (human) values and beliefs.
Of course one cannot exclude that visiting aliens might not want to get baptized, or convert to Judaism or Islam;
but, as ethnological records of European colonialism show, chances are high that current theocracies will be superseded by alien conceptions and theocracies.
The best terrestrial theology can hope for is the inclusion of their beliefs in a pantheon; just like the Romans processed peripheral deities.

For the aliens any exchange might be not so much a question of ``us attacking them'' (akin to aggressions of an ant toward an elephant);
nevertheless, a much subtler issue is related to the buildup of population pressure in subjugated territories; a kind of ``osmosis'' --
an ``imperial backflow'' of individuals from the periphery towards the centers, with negative effects on the advanced civilization~\cite{Raspail-cos}.
And unlike historic colonizations on Earth, any hypercivilization has very little to gain from conquering Earth, as human labor might
be considered excessively ineffective compared to the alien technology employed, and all material commodities can be plentifully
obtained and harvested elsewhere -- on uninhabited planets or rocks -- most likely in an automated, robotic way.
In short, from the point of view of a  galactic hypercivilization
Earth might be rather seen as a liability than an asset.
Aliens might perceive ``us'' as a potential threat or nuisance which needs to be contained.

Nevertheless, in the long run, contact with extraterrestrial, also intelligent, life forms may be unavoidable -- either by ``us'' discovering ``them'', or,
as has been argued earlier to
a lesser degree, buy ``them'' contacting ``us.''
Note in this context that the search for extraterrestrial intelligence by ``listening to signals in the skies''
appears utterly naive: it took human civilization a lapse of a century to, say, switch from amplitude \& frequency modulated signaling
(such as AM \& FM  radio and television communications and transmissions) to digital transmissions;
the latter one essentially being indiscriminative from white noise without keys to decipher the signal.
Conversely, {\it a posteriori}, white noise can be ``interpreted'' or ``deciphered'' as any (intentional) signal one can dream up:
in the most straightforward way consider a ``proper'' one-time pad (relative to both the intended message and
the supposedly random sequence extracted from white noise) matching suitably long
binary sequences from a white noise source; and  XOR'ing it, bit by bit, with the one time pad, such that a particular
bit of the cleartext is the sum modulo 2 of the corresponding bits of the sequences from white noise and the one-time pad, respectively.

Moreover, sequences which are, on the one hand,
provable random may, on the other hand, encode a wealth of facts.
For example,
Chaitin's halting probability for prefix-free programs $\Omega$
is both algorithmically incompressible and passes all statistical tests~\cite{martin-lof};
and yet it ``encodes the solutions to
all halting problems''~\cite{calude-dinneen06}.
Some cosmic signals may well transmit similar ``deep'' messages, but we would not be able to decipher them.

\subsection{Virtual realities}

At some point, we might have to accept that we are living in a virtual reality created for irritatingly trivial purposes, such as marketing~\cite{simula}.
The term {\em virtual reality}  needs further specification. Suppose that the virtual reality we inhabit is
endowed with, and at the same time limited to, universal Turing machine capacity; that is, it essentially is a paper machine.

There immediately appear to be two options (and a mixture thereof): either consciousness is an emergent property of paper machines,
as is claimed by the second Turing (hypo)thesis discussed earlier.
Or we are immersed in a participatory gaming environment so that whatever constitutes our consciousness ``runs''
on a substratum -- a ``beyond'' realm, possibly also a paper machine -- which is transcendent
relative to the gaming universe. In this latter, dualistic, scenario, communication is facilitated by an interface
allowing an information flow back and forth across the gaming reality and the ``beyond''.
In short, as stated in one of Godard's movies~\cite{godard-aa}, we are the ``dead on vacation''
(or a penitentiary, or  purgatory).

There is an important consequence of the second, dualistic, option:
notwithstanding Spinoza and Leibniz~\cite{sep-sufficient-reason}
anything transferred through the interface
lacks a  sufficient reason or cause in the respective other realm: the picture is incomplete if one just
concentrates on one such domain; both are tied together (``entangled'') through the interface -- but not causally so.

\subsection{Order from chaos}

Another possibility is that what we perceive as laws emerged from primordial chaos~\cite{hesiod+700-2};
and indeed all else~\cite{armstrong_1983,vanFraassen1989-VANLAS}, including paper machines, can be grounded in the latter.
To get a taste of this conception, Ramsey theory~\cite{GS-90} suggests that informally interpreted,
given arbitrary data, ``there cannot be no laws.''

The contemporary narrative of theoretical physics can be interpreted as corroboration of this assumption,
as, for instance, all photon emissions (spontaneous or stimulated) occur at random times.
As stated by Born in 1926~\cite[p.~866]{born-26-1} [English translation in \cite[p~54]{wheeler-Zurek:83},
{\em ``from the standpoint of our quantum mechanics, there is no quantity
which in any individual case causally fixes the consequence of the collision;
but also experimentally we have so far no reason to believe that there are some inner properties of the atom
which condition a definite outcome for the collision.
Ought we to hope later to discover such properties $\ldots$  and determine them in individual cases?
Or ought we to believe that the agreement of theory and experiment  --  as to the impossibility
of prescribing conditions? I myself am inclined  to give up determinism in the world of atoms.''}

\subsection{Relativity of morality}

The following could be understood as a gig into metaethics~\cite{sep-metaethics,fisher_2011},
a subject of concern already to Plato~\cite{plato-republic}, from a scientific angle.

\subsubsection{Means related uncooperative behavior due to brain injuries}

Let me point out up-front that, even before going into the problematic semantics of morality, there are ``trivial'' cases of ``sinful'' behaviors.
No ethical component whatever is involved insofar as the individual committing such behavior is concerned.
Because they originate from, and are caused by, such an individual's brain damages.

I am not talking about autonomous decisions which are willful in any form.
Any such cases do not at all relate to willful actions of an ``evil soul.''
This is about involuntary behavior --
 severe disorders and dysfunctionalities of the mind,
sometimes included in but not restricted to the {\em Diagnostic and Statistical Manual} (DSM-5)~\cite{DSM-5} --
and caused by a defective brain (functionality).
Even before religion, contemporary psychiatry is to blame for this confusion of ``profoundly immoral and wicked evil'' with sickness.
Let me quote Amen~\cite{Amen-TED-talk}:
{\em ``$\ldots$
psychiatrists are the only medical specialists that virtually never look at the organ they treat.
$\ldots$ just more medication thrown at him in the dark $\ldots$ or behavior therapy $\ldots$ which if you think about it is really cruel:
To put him on a behavior therapy program when behavior is really an expression of the problem [[but]] it's not the problem.''}
To consider such an individual ``sinful'' or ``evil'' is like blaming an immobilized person for not being able to walk.

Brain injury syndromes that express themselves by symptoms such as psychopathy appear to be widespread but often remain unrecognized.
Punishing a ``criminal'' for a disease of the brain is not only
inappropriate but also dangerous, as the untreated malady will reveal itself over and over again,
thereby causing harm to, and suffering of, the sick individual as well as others affected.
In such cases it appears to be utterly useless to call for morality and contemplate ``evil;''
all that is required is a cure or, if a cure appears to be unattainable by present means,
a containment of the causes (and effects) of such conditions.

\subsubsection{Historic and ethnic evidence}

Most religions come packaged or
``bundled up'' with their own moralities, as well as ``dos'' and ``don'ts'' --
such as the prohibition of cross-dressing in Deuteronomy 22:5 New International Version (NIV):
{\em ``a woman must not wear men's clothing, nor a man wear women's clothing, for the Lord your God detests anyone who does this.''}
-- it seems this applies to most contemporary women, and also to this Author.
And, by the way,   Deuteronomy 22:11 advises
{\em ``do not wear clothes of wool and linen woven together.''}
A little later one finds in
Deuteronomy 22:20 that {\em ``if $\ldots$ no proof of the young woman's virginity can be found,''}
Deuteronomy 22:21 {\em ``she shall be brought to the door of her father's house and there the men of her town shall stone her to death.''}
This Author finds it difficult to even attribute some allegorical~\cite{Maimonides1,Maimonides2} value to such verdicts
-- nonetheless, they are imparted in a core text of Abrahamitic tradition.

Indeed, from a contemporary European viewpoint, aspects of different religions, such as
the Vedic casts, partitioning humans into
{\it Brahmins} (theocracy),
{\it Kshatriyas} (administrators and warriors),
{\it Vaishyas} (artisans,  farmers),
{\it Shudras} (workers)
-- and not even mentioned but implicitly implied,
{\it Pariahs} (untouchables) --
present equally disagreeable clauses.
One might speculate and ethnology seems to suggest that,
as new, hitherto disentangled,  territories
learn about each other, their respective
moralities and ``customs'' oftentimes both ``overlap''
and are very different.
to quote Montaigne~\cite[Chapter~I, Section~23 , p~389.5-390.5/3320]{Montaigne-Essays},
{\em ``[C]~The laws of conscience which we say are born of Nature are born of custom; since man inwardly
venerates the opinions and the manners approved and received about him,
 he cannot without remorse free himself from them nor apply himself to them without self-approbation.
$\ldots$
[A]~But the principal activity of custom is so to seize us and to grip us in her claws
that it is hardly in our power to struggle free and to come back into ourselves,
where we can reason and argue about her ordinances.
Since we suck them in with our mothers' milk and since the face of the world is presented thus to our infant gaze,
it seems to us that we were really born with the property of continuing to act that way''}
It might be interesting to look for reasons of accord \& discord of the various customs and ethics encountered.

\subsubsection{Inconsistency of moralities}

A further troubling matter is that, even within a given ethnic and ethic framework,
there may exist two or more moral rules which appear right if seen individually, but are mutually contradictory if applied simultaneously.
Hence viewed relative to one rule -- which is thereby implicitly considered to be ``right'' -- all other such ``complementary'' ones
are found to be ``wrong;'' and {\it vice versa}.
The associated tasks to appropriate ``rightness'' and ``wrongness''
may by no means be trivial, and may depend on, and be relative to,
various priorities which cannot be weighted {\it a priory}.

One immediate, pressing example is the programming of autopilots~\cite{Wallach-Allen-MoralMachines} in cars {\it et cetera},
often referred to as the {\em trolley dilemma}~\cite{Foot1967-FOOTPO-2,Thomson-Trolley,Himmelreich2018},
and is related to collateral damage:
is it, for instance, right, to program the autopilot in a way
which would allow or even demand that, in a situation of exclusivity (exclusive or),
the life of two elderly people will be sacrificed to save one teenager; or should it be the other way round?
And in any case, how many elderly ought to compensate one teenager?
(A possible option would be to age identify potential targets individually,
and subsequently, weigh and maximize the sums of the average life expectancies of the respective target groups.)
And what about the passengers? Would anybody buy a car knowing that the autopilot might potentially sacrifice one's life for the sake of others?
Or should the autopilot decide whom to potentially kill and rescue on the basis of a random number generator?
Autonomously driving cars implicitly need built-in rules deciding such issues ``on the spot;''
that is, in cases of accidents or emergencies. Even if it does not want to decide, it has to act somehow
-- just as one cannot choose not to communicate~\cite{Watzlawick-1967}, or not find laws~\cite{GS-90,Calude2016}.

Another example is the appropriation of wealth among members of a society;
and, in particular, what
Dirac referred to as~\cite{dirac-81} {\em ``the basic principles of
modern human society. The first is contained in the fact that we
all acknowledge that it is a good thing for parents to take care of
and provide for their children, the second, that all children
should be given an equal chance, that is, the same opportunities,
for development.
$\ldots$
However, it takes little to see that these two principles,
each undoubtedly an excellent principle by itself, contradict each
other. For, if some parents make conditions better for their
children through sacrifices, these children will necessarily have
a better chance than those others whose parents either do not
bother or are unwilling to make comparable sacrifices.''}

A third example is a situation, exposed by Brecht's {\em The Good Person of Szechwan}~\cite{Brecht-GMvS}.
In this play, the lead character, the ``good person'' {\em Shen Teh},
in her kindness and goodness, becomes exploited by various characters of ``the mob'' until she becomes dysfunctional and broken.
At this point,
in order to  protect and sustain her good intentions, Shen Teh
has to impersonate as the created vicious cousin {\em Shui Ta} -- the ``bad guy.''

\subsubsection{Instability of societies}

The {M}atthew effect of accumulated (dis)advantages~\cite{merton-68}
-- ``the ones who have plenty will receive more, and the ones who have little will lose more'' --
is a built-in feature of our universe which fosters progress through catastrophes
but at the same time destabilizes societies, economies, and societies at large.
It is mentioned by Matthew 13:11-12 \& 25:29 as well as by Mark 4:25 and Luke 8:18 \& 19:26.

Mathematically, the {M}atthew effect is formalized by the compound interest,
which grows exponentially.
The growth or decline appears linear at  small
timescales: at ``small enough'' deviations from its point of Taylor expansion an exponential function can
be approximated by a linear Taylor polynomial; thereby neglecting the effects of degree two and higher.

Compound interest is at work in all aspects of human life and societies, eventually yielding instabilities due
to the buildup of huge imbalances.
Negating its devastating -- but also revolutionary and reviving -- thrust is akin to ``running against a wall.''

\subsubsection{Paradoxes of freedom, tolerance, and weakness}

According to Plato~\cite[562b-563e]{plato-republic},
an excess of liberty and freedom yields anarchy and then tyranny.
Because if (collections of) individuals consider themselves free of any rules,
the rules and authority get corrupted and ``inverted''~\cite[562e-563a, p.~275]{plato-republic}: {\em ``A father, for example, gets used to being like a child,
and being afraid of his sons.
A son gets used to being like his father.
He feels no respect and fear of his parents. All he wants is to be free.
Immigrants are put on par with citizens, and citizens with immigrants.''}
Chaos ensues~\cite[563d-e, p.~276]{plato-republic}:
{\em ``In the end $\ldots$ they take no notice of the laws $\ldots$ in their determination
that no one shall be the master over them in any way at all.''}
Then, according to Plato, these squanderings and excesses of freedom~\cite[563e-564a, p.~277]{plato-republic}: {\em ``produce a violent reaction
in the opposite direction. $\ldots$ the chances are that democracy is the ideal place to find the origin of tyranny
-- the harshest and most complete slavery arising $\ldots$ from the most extreme freedom.''}

Popper phrases the {\em paradox of tolerance}~\cite[Chapter~7, Footnote~4]{Popper-TOSAIE}
in terms of a ``diagonalization argument''
(a kind of paradoxical self-negation by substitution very common in metamathematics~\cite{smullyan-92,Smullyan1993-SMURTF,book:486992}):
{\em ``Unlimited tolerance must lead to the disappearance of tolerance.
If we extend unlimited tolerance even to those who are intolerant,
if we are not prepared to defend a tolerant society against the onslaught of the intolerant,
then the tolerant will be destroyed, and tolerance with them.''}
(I disagree with Popper's interpretation of Plato's paradox of freedom reviewed earlier.)

One strategy, inspired by metamathematics~\cite{Smullyan1993-SMURTF},
is to avoid such paradoxes by {\em restricting} the respective means -- in this case freedom and tolerance --
to instantiations which cannot produce an undesired event (such as inconsistencies).
Of course, the question of extent and appropriateness of such measures of censorship remains unresolved
and will be discussed later.

Nietzsche~\cite{Nietzsche-GM} set out to criticise the kind of ``slave morality''
he ascribed to Christianity  (and, one might argue, by transitivity, socialism, and communism)
which he considered being based upon
the re-interpretation
--
in an Orwellian {\em newspeak} sense, {\em doublethink}~\cite{Orwell-1984}
--
of weakness as strength; and strength as a weakness.
(This can be considered just another kind of diagonalization.)
By contrast, ``master morality'' and the ``will to power''
as it is exposed in Thucydides'
{\em History of the Peloponnesian War}, in particular,
the {\em Melian dialogue}~\cite[Chapter~V, \S~89]{Thucydides-Mynott}
taking place in the summer of the sixteenth war year,
appears pragmatic and sober:
{\em ``in the human sphere judgments about justice are relevant only
between those with an equal power to enforce it, and that the possibilities
are defined by what the strong do and the weak accept.''}
{\em ``$\ldots$ submission would save you from suffering a most
terrible fate, while we would profit from not destroying you~\cite[Chapter~V, \S~93]{Thucydides-Mynott}.''}
{\em ``$\ldots$ your enmity does us less harm than your
friendship; that would be taken by our subjects as a sign of weakness on
our part, while your hatred is a sign of our strength.~\cite[Chapter~V, \S~95]{Thucydides-Mynott}.''}

\subsubsection{Morality by game theory}

From what has been mentioned earlier it should be clear that, at least by present rational means,
the distinction between ``right'' and ``wrong,'' or ``good'' and ``bad'' behavior cannot
be given in any objective, absolute sense.
(This cannot outrightly exclude absolute or objective moral criteria but,
because of possible paradoxa from constructions involving self-contradicting substitutions,
their prospects can be conjectured to be slim.)
Moreover, from a scientific point of view,
many commandments, such as Moses'
{\em ``thou shalt not kill,''}  or {\em ''thou shalt not steal''}
are sociologically, psychologically and politically advantageous.
But even these profound desiderata appear means and context relative;
inhibited and even reversed, for instance in times of war.
For the sake of an example,
consider McNamara
referring to the  firebombing air raids of Japanese cities during World War II~\cite{Morris-fog-of-war}:
{\em ``was there a rule $\cdots$ that  said you shouldn't bomb, shouldn't kill, shouldn't
burn to death a hundred thousand civilians a night?
LeMay said if we lost the war we would all have been prosecuted as war criminals.
And I think he's right. He and I'd say I, were behaving as war criminals.''}

Oftentimes food restrictions -- in particular the killing of animals and the processing of their cadavers --
have been and still are medically reasonable; in particular in hot countries.
Others commandments, such as the ones discussed earlier in Deuteronomy 22,
or ritualistic ceremonies,
might be considered as expressions of intolerance, or even malignant
Obsessive-Compulsive Disorders at worst.
Many such traditions and rules appear inconsistent, outdated and queer.

Most importantly
the entire body of ethics is incapable of handling
{\em quantitative} issues of appropriation and guidance.
These involve central questions of the (re)distribution of wealth, but also ways to steer vehicles.
Ancient traditional moralities are therefore insufficient for the need for technologically advanced civilizations.

Therefore it has been proposed to base a quantitative, formalized morality
on mathematical concepts; in particular,
on game theory~\cite{Braithwaite-gt,chamberlin_1989,Nowak:1995:AMH,Verbeek-2002,Cavagnetto-14,sep-experimental-moral,sep-game-ethics}.
Instead of going into details consider a widely successful
{\it TIT FOR TAT} strategy~\cite{Axelrod-80a,Axelrod-80b,Nowak:1995:AMH}:
{\em ``cooperate initially, and thereafter cooperate if the other
side cooperated last time and defect if the other side defected last time.''}
Computer simulations (such as for the iterated Prisoner's Dilemma) and mathematical analysis demonstrate
that cooperation and altruism based upon reciprocity
emerges and proves stable in a world without a central authority,
and inhabited by egoists --
an evolution of cooperation and secular norms~\cite{Axelrod1390,Axelrod1385,Axelrod-85,axelrod_1986}.

In such a relative, emergent, view morality presents itself as a huge and complex canvas of partially overlapping and partially conflicting strategies
at various levels;
like an onion of unknown extension and depth,
a patchwork of fined tuned compromises.

Nietzsche could not have foreseen these developments.
Therefore his rant against religion, in particular,
Christendom, was strongest in criticism but weakest when it came to positive alternatives
-- for instance, the ``will to power'' is all but one of many principles upon which good strategies need to be based.

\section{Deconstruction of the scientific claim of truth}

So far it appears that science has been on the offensive and religion on the defensive.
Nothing could be farther from the subject in question.
Because all matters discussed earlier barely scratched the surface, the phenomenology,
of our existence, and the existence of the universe ``around us.''

Alas, as all matters invented, created and practiced by humans
-- and despite its liberating and beneficial consequences --
science itself cannot claim any  absolute truth or exclusivity but remains
{\em ``suspended in free thought.''}
By its own skeptical standards, nothing indicates that its very basis is not formed by metaphysical concepts
grounded solely in our beliefs in them, which are further expanded into narratives of great ``material''
and practical usefulness
--
resembling songlines or dreaming tracks,
the songs of Dreamtime,
guiding indigenous Australians through their territories.
Idealism has simililar suspicions by claiming that~\cite{Goldschmidt2017-idealism-Ch3}
{\em ``the world is mental through-and-through.''}

To quote Nietzsche again~\cite{Nietzsche-WahrheitLuege,NietzscheReader},
{\em ``What then is truth? A movable host of metaphors, metonymies, and anthropomorphisms:
in short, a sum of human relations which have been poetically and rhetorically
intensified, transferred, and embellished, and which, after long usage, seem to a
people to be fixed, canonical, and binding. Truths are illusions which we have forgotten
are illusions~$\dots$ After all, what is
a law of nature as such for us? We are not acquainted with it in itself, but only with
its effects, which means in its relation to other laws of nature -- which, in turn, are
known to us only as sums of relations. Therefore all these relations always refer again
to others and are thoroughly incomprehensible to us in their essence~$\dots$
All that conformity to law, which impresses us so much in the movement
of the stars and in chemical processes, coincides at bottom with those properties which
we bring to things.''}
Similar thoughts have been expressed by Camus~\cite{camus-mos,camus-mose}:
{\em ``You explain
this world to me with an image. I realize then that you
have been reduced to poetry: I shall never know. Have I
the time to become indignant? You have already changed
theories. So that science that was to teach me everything
ends up in a hypothesis, that lucidity founders in metaphor,
that uncertainty is resolved in a work of art. What need
had I of so many efforts? The soft lines of these hills and
the hand of evening on this troubled heart teach me much
more.''}

In a less poetic and more analytic style,
many philosophers of science have expressed similar thoughts,
in particular also Lakatos~\cite{lakatosch}
and van Fraassen~\cite{vanFraassen1989-VANLAS}.
Hertz~\cite[Introduction]{hertz-94,hertz-94e} expressed it this way:
{\em ``We form for ourselves images [[chimera, the German original is {\em Scheinbild}]] or symbols of external objects;
and the form which we give them is such that the necessary
consequences of the images in thought are always the images of
the necessary consequents in nature of the things pictured.
$\ldots$~we do not
know, nor have we any means of knowing, whether our conceptions
of things are in conformity with [[the things]] in any other
than this one fundamental respect.''}
Einstein later notes~\cite{Einstein-34}:
{\em ``Reason gives the
structure to the system; the data of experience and their mutual
relations are to correspond exactly to consequences in the theory.
On the possibility alone of such a correspondence rests the value
and the justification of the whole system, and especially of its
fundamental concepts and basic laws. But for this, these latter
would simply be free inventions of the human mind which admit
of no a priori justification either through the nature of the
human mind or in any other way at all.''}
However, although acknowledging such conceptual issues,
Einstein was no idealist and strongly believed in the possibility of such a correspondence
but gives no reasons why this should be so.

What then are some aspects of the scientific narratives or songlines presently told?
That ``almost all'' (formally of Lebesgue measure one)  space is totally empty, a void of nothingness.
Immersed in that void are point particles of zero extension,
as enumerated by the {\em Review of Particle Physics}~\cite{PhysRevD.98.030001}.
These ``bricks'' forming all objects can be grouped into two types:
so-called bosons and fermions.
Whereas the former ones like to ``clog together'' if they are identical,
the latter ones abhor their kin: there cannot be two identical ones at the same time and place,
thus forcing an ``extension'' of fermions such that they ``feel apart comfortably.''
All elementary bosons but ones mediate the forces between fermions,
and are therefore responsible for repulsion and attraction,
and also for the formation of intermediate structures called ``atom'' (a misnomer),
whose nucleus also consist of fermions, which in turn consist of fermions and bosons.
One boson, the so-called Higgs particle, mediates masses and the different strength of interactions.

Every individual process, in particular, the emission and absorption of light occurs at instances which are irreducibly ``random;''
that is, unpredictable by any paper machine. There is no sufficient cause for such emissions and absorption;
particles are emitted by {\em creatio continua}.  What you see is from spontaneous emissions which can only be
predicted probabilistically but not individually.

So, according to the present scientific narrative,
swimming in the waters of Macedonia's Lake Ohrid, or of the Irrsee in Austria's Salzkammergut,
is moving through an emptiness, nothingness --
swimming is traversing a void containing particles which,
if one would attempt to measure their size,
have no extension at all
-- for all practical purposes they behave like singular points.
And your body is also mostly emptiness, nothingness, a void, just like all solid or liquid objects.
Those two voids interact
-- and their fermions inhabiting them abhor being together at the same place at the same time --
and allow you to swim.
Thereby, the experience of an object ``feeling hard'' and impenetrable
is reduced to the interaction between this object and another one
which it ``touches'' (for instance your hand),
and thereby cannot penetrate it.

Never mind the huge epistemic gaps in our comprehension of the universe:
the forces discovered so far have only been partially identified and unified into a comprehensive ``standard theory''
-- alas defying gravity.
There are issues related to faraway things in the sky -- such as the rotation of galaxies --
which are inexplicable by the aforementioned songlines.
In order to cope with these deficiencies, a hypothetical form of matter has been postulated and
assuringly called ``dark matter'' (as it does not emit or interact with light).
It is supposed to be ubiquitous in the universe and amounting for most, that is, more than 80{\%},
of the stuff (matter) there exists.
The standard model of cosmology also requires another hitherto unknown form of energy known as ``dark energy''
permeating all of space and accelerating the expansion of the universe.
Dark energy and dark matter combined are supposed to account for most, that is, 95{\%}, of the total
energy of the universe.
Current songlines also contend that the universe started with a ``big bang'' and expands ever since.

Quantum mechanics basically amounts to a theory of vectors and their generalized length preserving rotations.
It is inconsistent -- in postulating irreversible
measurements somehow arising from a ubiquitous reversible state evolution~\cite{v-neumann-49,v-neumann-55,everett,schroedinger-interpretation}
--
and nevertheless highly successful (just as Cantorian set theory) for all practical purposes~\cite{bell-a},
predictions and guidance.

\section{Hierarchies of ``throwaway'' entities expressing lower levels of description}

Before moving on to the metaphysics of existence let me amend aspects which come up in the discussion of evolution:
that~\cite{Hamilton-1963} {\em ``the ultimate criterion
which determines whether [[a gene]] will spread is not whether the behavior
is to the benefit of the behaver but whether it is to the benefit of the gene''}
--
``selfish'' genes~\cite{dawkins-selfishgene}
express themselves through individuals~\cite{dawkins-video}:
{\em ``an individual organism is a throwaway survival machine
for the self-replicating coded information which it contains.''}

Alas this is an example of
not seeing the wood for the trees, as, to paraphrase Dawkins,
{\em genes represent throwaway survival machines for the laws of the universe which they  express.}
Rather than dealing with just two layers of description
one should take into account a much wider picture,
thereby including a multiply layered structure of emergent entities,
each layer having its own justification and characteristic~\cite{anderson:73}:
the laws of the universe could be perceived as the expressions of the universe and of its existence (some would call this creation)
--
the formation of genes could in turn be perceived as the expression
of the laws governing the universe
--
the genes themselves express themselves in the individual bodies
--
and individual minds  could in turn be perceived as the expression
of the bodies they are associated with.

In this view emergence continues starting from a possibly random, unorganized and ``elementary''
layer of existence~\cite{Exner-1908,Stoeltzner-1999,svozil-2018-was}.
The process of (self-)organizatzion
proceeds to the ``formation of laws and matter,'' and towards
(by the strong second Turing hypothesis) the formation of conscious minds capeable to reflect upon the situation.

Therefore, we are inclined to maintain that this hierarchy of apparently highly organized patterns and structures,
like an iceberg which is only visible above sea level,
is ultimately grounded in the great metaphysical abyss of existence.
This issue will be discussed next.

\section{Metaphysics 101: the enigma of existence}

Previous lives were ``full of wonders, mysticism and miracles''~\cite{stace-map,Cotter-1999,Gerolemou-18}
which have ceased to occur for various reasons; also because of our scientific capacity
to causally explain the phenomena.
Even if we don't know any causes we tend to assume that this is an epistemic issue,
and causes exist: we just don't know them.
To quote a psychoanalytic motto, {\em ``where id was, there shall ego be.''}

Is the world devoid of any even indirect instantiation of transcendence?
Maybe there exist human minds who do not find their experience of existence
to require any explanation whatsoever: they (and by transgression to an outside world,
the universe they live in) exist; and that's about it.
Indeed, the enigma of existence might be considered a (sometimes lyrically decorated~\cite{Hahn1930})
{\em pseudo-statement}~\cite{Hume-Enquiry,Carnap1931,Carnap-1931-engl}.
This critique is
based on the assumption that it is possible to find an ``archimedian ontological anchor (or handle),''
such as means and mind independent empirical or logical criteria,
--
that is, in Hume's words~\cite[Sect.~XII[34], p.~120]{Hume-Enquiry},
any abstract reasoning concerning quantity or number, or experimental reasoning concerning matter of fact and existence
--
which could serve as a solid foundation of judgements and progress.
Indeed it might come as no surprise that
Carnap~\cite{Carnap1931,Carnap-1931-engl} cites Hilbert~\cite{Hilbert1931};
as both publications had not yet absorbed the means relativizing impact of G\"odel's findings~\cite{godel1} for metamathematics
which appeared in that same year 1931.
However, in what follows it is suggested, or rather maintained or assumed,
that any such critical position, albeit conceivable, cannot be sustained by any conscious entity ``in the long run.''
(I am not talking about the short-term suppression of these issues for various reasons such as pragmatism or
avoidance of anxieties and bewilderments.)
On the contrary, it will be argued that the mind-boggling fact of our existence is a subjective experience of metaphysical nature
\begin{itemize}
\item[(i)]
immune to sobering science and rational thought;
\item[(ii)]
in its most individualistic form, it is even immune to solipsism as well as idealism~\cite{stace,Goldschmidt2017-idealism}.
Because whoever acknowledges one's own existence like Descarte's {\it ``cogito, ergo sum''}
can at the same time contemplate about the enigma of existence
``why do I exist rather than not exist?''
This quest is independent of whether one is alone, in a virtual environment of a  participatory game,
in a multiple layered simulation~\cite{svozil-93,svozil-nat-acad,Bostrom-sim}
or in exchange with other autonomous individuals;
\item[(iii)]
 in its irreducible incomprehensibility,
presents a clear indication and corroboration for a mindset wisened by the Socratic paradox ``I know that I don't know''
(also used~\cite{Rumsfeld-2001} for waging war);
\item[(iv)]
by analogy suggests that just because we do not understand a thing or two (and maybe never will)
those entities cannot exist: existence is both incomprehensible and existent.
\end{itemize}

Existence is at the metaphysical root of everything.
And everything which can be scientifically asked or searched contains in its deepest layer also this enigma of existence.
To quote Heidegger~\cite[Chapter~1,\S~1, p.~5/37]{Heidegger-1935,Heidegger-2000},
{\em ``our question $\ldots$ is necessarily asked, knowingly or not, along with every question.''}

The enigma of existence (in a nonsolipsistic form) has many fathers and ancient
roots~\cite{sep-nothingness,Rundle-04,gericke2008,gruenbaum-2008,Krauss-2012,Lynds-2012,Bilimoria2012,Goldschmidt2013-GOLTPO-33,carol-witsrtn-2018}.
After Leibnitz introduced~\cite[p.~639]{Leibniz1989}
{\em ``the principle that nothing takes place without a sufficient reason
$\ldots$ the first question which we have a right to ask will be, `Why is
there something rather than nothing?' For nothing is simpler and easier than something.''}
Later Wittgenstein stated~\cite[6.44]{Wittgenstein-TLP}
{\em ``Not {\em how} the world is that is mystical, but {\em that} it exists.''}
Heidegger posed the {\it Angstfrage}
in his Freiburg lectures on metaphysics~\cite{Heidegger-1929,Heidegger-1935}:
{\em the fundamental question of
metaphysics $\ldots$: why is there something [[or that which exists]] rather than nothing?}
(The German original
{\em ``die Grundfrage der Metaphysik $\ldots$: Warum ist \"uberhaupt Seiendes und nicht vielmehr Nichts?''}
has been inadequately translated into English~\cite{heidegger_krell_1998,Heidegger-2000}.)

Of course, as has been mentioned earlier, an immediate response might be that metaphysical questions such as the aforementioned
enigma of existence -- ``why is there something rather than nothing?'' --
are meaningless,
as only empirical scientific knowledge derived from experience
or the analysis of language with logic~\cite{Hume-Enquiry,Hahn1930,Carnap1931,Carnap-1931-engl} is meaningful.
To quote Wittgenstein~\cite[4.11]{Wittgenstein-TLP},
{\em ``the totality of true propositions is the whole of natural science (or the whole corpus of the natural sciences).''}
Wittgenstein's final slogan~\cite[7]{Wittgenstein-TLP}
{\em ``What we cannot speak about we must pass over in silence''}
serves a sort of cartoon blog (this author is bewildered by the fame Wittgenstein received outside of Vienna --
 there he might have been fared as  not very original; a talkative narcissist), and puts him in a row with
venerable representatives of the ``why botherers''~\cite{bell-a} such as Dirac and Feynman~\cite{feynman-law,clauser-talkvie}.
In any case, nagging questions should be kept in mind with {\em evenly-suspended attention}~\cite{Freud-1912,Freud-1912-e}
rather than suppressed, as any kind of suppression (into the unconsciousness) alleviates oneself from coping
with the issue immediately but bears the danger of neurosis (and resurrection of that which has been suppressed) by integration
of the suppressed content into one's character -- in this case, by analogy, into
the realm of human thought.
Also, disallowing questions might turn out to inhibit innovation.

Another response would be a subjective one by meditating  about
one's own undeniable experience of existence.
Thereby, one could ask about the enigma of one's own existence,
and acknowledge its incomprehensibility.

If one accepts the issue as relevant then it is robustly so with respect to all variants of attacks.
For instance, we might be living on a ``Russian doll'' like layered virtual reality, even without any ``bottom layer;''
and we may even get a feel for why these structures have been made by ``looking down''
and acknowledging one transcendent layer.
But then the question remains about the existence
of the respective ``lower'' layer, until by transitivity, one has reached the ``bottom layer'' or continues asking.

Allowing the quest for the enigma of existence opens up an abundance of possibilities and narratives -- often nonverifiable or unfalsifiable
-- which could lead one to a better, more humble, understanding of our limits of thought,
and of that which could be conceived as being possible,  but need not necessarily be so.
This suggests a more open-minded, lenient interpretation of the scriptures;
and of religious experience in general.

Let me close with a pagan {\it adagium} cited by Erasmus of Rotterdam~\cite[3:1232, p.~240-241]{Erasmus-Adagia1001-1500}:
{\em VOCATVS ATQVE INVOCATVS [[or NON INVOCATVS]] DEVS ADERIT:
called or not called, God will be present.}


\begin{acknowledgments}
I kindly acknowledge enlightening discussions with, and suggestions by, Silvia Jonas and Irem Kurtsal.
All misconceptions and errors are mine.
\end{acknowledgments}

\onecolumngrid


%
\end{document}